\documentclass[prl,twocolumn,showpacs,floatfix,superscriptaddress,amsmath,
amssymb]{revtex4}

\usepackage{graphicx}
\usepackage{dcolumn}
\usepackage{bm}
\usepackage{pstricks}

\def\nn{\nonumber}
\def\va1{\vec{a}_{1}}

\def\vb1{\vec{b}_{1}}

\def\vd1{\vec{\delta}_{1}}

\newcommand{\ba}{\begin{eqnarray}}
\newcommand{\ea}{\end{eqnarray}}
\def\be{\begin{equation}}
\def\ee{\end{equation}}

\def\vx{\vec{x}}
\def\vy{\vec{y}}
\def\vra{\vec{r}_{\alpha}}
\def\vrb{\vec{r}_{\beta}}

\begin{document}

\title{Quantum disordered phase on the frustrated honeycomb lattice}

\author{ D.C.\ Cabra }
\affiliation{Departamento de F\'isica, Universidad Nacional de La Plata,
C.C.\ 67, 1900 La
Plata, Argentina.}
\affiliation{Facultad de
Ingenier\'\i a, Universidad Nacional de Lomas de Zamora, Cno.\ de Cintura y Juan
XXIII, (1832)
Lomas de Zamora, Argentina.}
\affiliation{Institut de Physique et Chimie des Mat\'eriaux de Strasbourg, UMR7504,
CNRS-UdS, 23 rue du Loess, BP43, 67034 Strasbourg Cedex 2, France.}

\author{ C.A.\ Lamas}
\affiliation{Departamento de F\'isica, Universidad Nacional de La Plata,
C.C.\ 67, 1900 La
Plata, Argentina.}

\author{ H.D.\ Rosales}
\affiliation{Departamento de F\'isica, Universidad Nacional de La Plata,
C.C.\ 67, 1900 La
Plata, Argentina.}

\begin{abstract}
In the present paper we study the phase diagram of the Heisenberg model on the
honeycomb lattice with antiferromagnetic interactions up to third neighbors
along the line $J_2=J_3$ that include the point $J_2=J_3=J_1/2$, corresponding to the highly frustrated point where the classical ground state has macroscopic degeneracy.
Using the Linear Spin-Wave, Schwinger boson technique followed by a mean field decoupling and
exact diagonalization for small systems we find an intermediate phase with a spin gap and
short range N\'eel correlations in the strong quantum limit $(S=\frac12)$.
All techniques provide consistent results which allow us to predict the
existence of a quantum disordered phase, which may have been observed in recent
high-field ESR measurements in manganites.
\end{abstract}

\maketitle


The two-dimensional Heisenberg model in frustrated geometries has received a lot of attention in the
last years \cite{Sachdev1,Sachdev2,Sandvik,Poilblanc}. One of the reasons for this
interest is the common belief that geometrical frustration in two-dimensional
(2D) antiferromagnets may enhance the effect of quantum spin fluctuations and hence suppress
magnetic order giving rise to a spin liquid \cite{Anderson}.
One candidate to test these ideas is the honeycomb lattice, which is bipartite
and has a classical N\'eel ground state, but due to the small coordination number $(z=3)$,
quantum fluctuations could be expected to be stronger than those in the square
lattice and may destroy the antiferromagnetic long-range order (LRO).

The study of frustrated quantum magnets on the honeycomb lattice
has also experimental motivations: on the one hand, recent ESR
experiments in high magnetic field on Bi$_3$Mn$_4$O$_12$(NO$_3$),
which is described by a $S=3/2$ honeycomb Heisenberg model, have lead to the conjecture that
geometric frustration plays an important role in removing the long range magnetic order \cite{ESR}.
On the other hand, the family of compounds $BaM_2(XO_4)_2$
with $M=Co,Ni$ and $X=P,As$, consists of magnetics
ions $M$ arranged in weakly coupled frustrated honeycomb lattices with spin $S=1/2$ for Co
and $S=1$ for Ni \cite{libro_experimental}.

Last but not least, new
possibilities to design interacting lattice systems with controlled geometry
emerge: modern strategies in chemistry
open a route to synthesize new materials with a desired lattice structure and
intersite interaction \cite{design} and the controlled setup of optical lattices
for cold atoms would allow to create arbitrary lattice structures as
well as to tune the interactions \cite{cold_atom_a,cold_atoms}.

In the present paper we study the Heisenberg model on the honeycomb lattice
with first ($J_1$), second ($J_2$) and
third ($J_3$) neighbors couplings \cite{Fouet},
along the special line $J_{2}=J_3$.
Using Linear Spin-Wave theory, the Schwinger boson technique
and exact diagonalization we find strong evidence for the existence of an intermediate disordered region
where a spin gap opens and spin-spin correlations decay exponentially. Although our results correspond
to a specific line, we conjecture that the quantum disordered phase that we have found in the vicinity of the 
tricritical point extends in a finite region around it.

\begin{center}
\begin{figure}[t]
\includegraphics[width=0.42\textwidth]{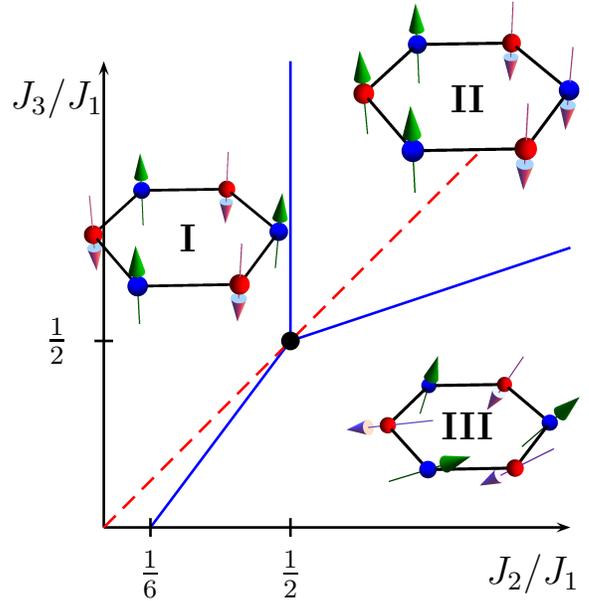}
\caption{(Color online) Classical phase diagram for the model (\ref{eq:Hspin_general}). There are three different phases schematized by the cartoons: The N\'eel phase (I)
 has two (antiparallel) ferromagnetic triangular sublattices (blue and red balls), the collinear phase (II) where ferromagnetic chains are antiparallel and the spiral phase (III). We focus on the dashed line corresponding to $J_2=J_3$ that includes the special point ($J_2=J_3=J_1/2$) where the ground state is infinitely degenerate.}
\label{fig:classic}
\end{figure}
\end{center}
The Heisenberg model on the honeycomb lattice is described by
\small
\ba
 \label{eq:Hspin_general}
 H =J_1\sum_{NN} \vec{\bf{S}}_{i}\cdot
\vec{\bf{S}}_{j} +J_2\sum_{NNN} \vec{\bf{S}}_{i}\cdot
\vec{\bf{S}}_{j}+J_3\sum_{NNNN} \vec{\bf{S}}_{i}\cdot
\vec{\bf{S}}_{j}.
\ea
\normalsize
were, $\vec{S}_{i}$ is the spin operator on site $i$.

The classical model \cite{Rastelli,Fouet}
displays different zero temperature phases shown in Fig.\ (\ref{fig:classic})
with a tricritical point at $J_2=J_3=\frac12 J_1$. At this point, the Hamiltonian can be written,
up to a constant term, as a sum over edge-sharing plaquettes
\ba
H=\frac{J_1}{2}\sum_{\gamma}|\vec{\bf{S}}_{\gamma}|^{2} ,
\ea
where $\vec{\bf{S}}_{\gamma}=\sum_{l\in \gamma}\vec{\bf{S}}_{l}$ is
the total spin in the hexagon $\gamma$.

At this particular point any state with vanishing total spin per
elementary hexagon is a classical ground state, giving rise to a
large GS degeneracy. This kind of situation is reminiscent
of what happens in the $J_1-J_2$ square lattice case for
$J_2/J_1 = 0.5$ \cite{square}, which presents a disordered
spin gapped phase around this point at the quantum level.

Motivated by this analogy, we study here the region around the
tricritical point $J_2=J_3=\frac{1}{2}J_1$ keeping $J_2=J_3$
as shown in Fig.\ \ref{fig:classic}.


We construct in the following the corresponding magnetic phase diagram
with the results obtained by using spin-waves, Schwinger bosons and ED
techniques.

\begin{figure}[t]
\includegraphics[width=0.44\textwidth]{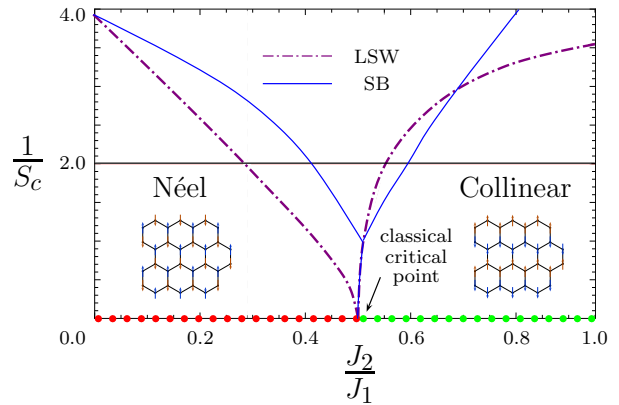}
\caption{(Color online) Inverse of the critical pin $S_c$ as a function of $J_2$:
The dashed-point line shows the bound obtained by LSWT. Above this
line, quantum fluctuations destroy the magnetic order, the staggered magnetization go
to zero. For $S=1/2$ (full black line), there is a small range $0.29\lesssim
J_2/J_1\lesssim0.55$ where there is not magnetic order; the full blue-line
corresponds
to the edge obtained by SBMFT. For the
$S=1/2$ case, there is a range $0.41\lesssim J_2/J_1\lesssim0.6$ where the
system has a spin-gap indicating a quantum disordered phase (see Fig.
\ref{fig:gap800}). The dotted-line correspond to the classical limit
$S\to\infty$ where the ground state correspond to
the Neel phase in the region $J_2/J_1<0.5$ ,
while for
$J_2/J_1>0.5$ the ground state correspond to the CAF phase.}
\label{fig:Sc_NAF}
\end{figure}

The classical phase diagram reduces to that shown in the line $1/S_{c} = 0$ of Fig \ref{fig:Sc_NAF}
where two collinear phases meet at the classical critical point $J_2/J_1 =0.5$.
A general ordered planar spin configuration can be written as:
\ba
{\vec{S}}_{{\vec{R}},\alpha}=
  S \big( \cos\left({\vec{Q}}\cdot {\vec{R}}+\phi_{\alpha}\right) { \check{e}_1}
+   \sin\left(\vec{Q}\cdot \vec{R}+\phi_{\alpha}\right) {\check{e}_2}
   \big)
\ea
where $\vec{R}$ is the position of the unit cell and  $\check{e}_1,\check{e}_2$
are the primitive vectors of the direct
lattice. The magnetic configuration is characterized by  the ordering
wavevector $\vec{Q}$
and the internal phase $\phi_{\alpha}$, where $\alpha=A,B$ is an internal index
on each sublattice. There are two different phases for the
classical ground state: for
$J_2/J_1<0.5$ the ground state corresponds to the N\'eel phase, with ordering
vector $\vec{Q} =(0,0)$ and $\phi_{A}-\phi_{B}=\pi$ and for
$J_2/J_1>0.5$ the ground state corresponds to a collinear phase with
$\vec{Q}=(2\pi/\sqrt{3},0)$  and $\phi_{A}-\phi_{B}=\pi$. In the following we
call this phase columnar antiferromagnetic (CAF) phase.
Using linear spin-wave theory (LSWT) \cite{Auerbach,aclaracion}, we have investigated the
stability of the classical phase diagram. The results obtained are summarized
in Fig.\ \ref{fig:Sc_NAF} as a function of the spin-$S$ and
the frustration $(J_2/J_1)$ (we do not present details in the
different steps of LSWT  since these are standard
\cite{Auerbach,Spinwave1,Spinwave2}). The edge of the ordered phases was obtained by
finding the frustration $(J_2/J_1)$
at which the quantum fluctuations destroy the classical order, {\it i.e.} where
the order parameter vanishes.
The results show that quantum fluctuations reduce the N\'eel stability
around the classical point
$J_2/J_1=1/2$. For $S=1/2$, N\'eel order is found for $0<J_2/J_1\lesssim 0.29$
and the CAF order for $0.55\lesssim J_2/J_1$.
To further analyze the region $0.29\lesssim
J_2/J_1\lesssim 0.55$ which is seen as a non-magnetic phase within LSWT we
now use the Schwinger-boson mean field theory (SBMFT) and Exact Diagonalization.

The Heisenberg model on the honeycomb lattice
was studied using SBMFT by Matson et al \cite{Matson} for antiferromagnetic
interactions at first and second neighbors.
Here we study Hamiltonian (\ref{eq:Hspin_general}) using a rotationally
invariant version of this technique, which has proven successful in incorporating quantum
fluctuations \cite{Trumper1,Trumper2,Coleman}.

In the Schwinger-boson approach,
the Heisenberg interaction is
written as a biquadratic form. In this representation the spin
operators on each site are replaced by two species of bosons via
the relation:
\ba
 \vec{\mathbf{S}}_{\vx}=\frac{1}{2}\vec{\mathbf{b}}^{\dag}_{\vx}.\vec{\sigma}.\vec{\mathbf{b}}_{\vx},
\ea
 where ${\vec{\bf b}_{\vx}^{\dagger }}\!=\!({\bf b}_{\vx\uparrow }^{\dagger },{\bf b}_{\vx\downarrow }^{\dagger })$ is a bosonic spinor, $\vec{\sigma}$ is the vector of
Pauli matrices, and there is a boson-number restriction $\sum_\sigma
\mathbf{b}^{\dag}_{\vx\,\sigma}\mathbf{b}_{\vx\,\sigma}\!=\!2S$ on each site.

 With this representation, the rotational invariant spin-spin
interaction can be written as
\ba
  \vec{\mathbf{S}}_{\vec{x}}\cdot
\vec{\mathbf{S}}_{\vec{y}}=:\mathbf{B}^{\dag}_{\vx,\vy}
\mathbf{B}_{\vx,\vy}:-\mathbf{A}^{\dag}_{\vx,\vy} \mathbf{A}_{\vx,\vy}
\ea
 where $\sigma=\pm$,  $:\mathbf{O}:$ indicates the normal
ordering of the operator $\mathbf{O}$ and we define the $SU(2)$
invariants
\ba
 \mathbf{A}_{\vx,\vy}&=&\frac12 \sum_{\sigma} \sigma
\mathbf{b}_{\vx,\sigma}\mathbf{b}_{\vy,-\sigma}\\
 \mathbf{B}^{\dag}_{\vx,\vy}&=&\frac12 \sum_{\sigma}
\mathbf{b}^{\dag}_{\vx,\sigma}\mathbf{b}_{\vy,\sigma}.
\ea
The operator $\mathbf{A}_{\vx,\vy}$ creates a spin
singlet pair between sites $\vx$ and $\vy$ and
$\mathbf{B}_{\vx,\vy}$ creates a ferromagnetic bond, which implies the
intersite coherent hopping of the Schwinger bosons.
This rotational invariant decoupling, which enables to treat
ferromagnetism and antiferromagnetism on equal footing, was introduced by
Ceccato et al \cite{Trumper1}. Later, Flint and
Coleman \cite{Coleman} presented a generalization to large N that shows
substantial improvements over the $Sp(N)$ approach.

To generate a mean field theory, we perform the Hartree-Fock decoupling
\begin{widetext}
\ba (\vec{\mathbf{S}}_{\vx+\vra}\cdot \vec{\mathbf{S}}_{\vy+\vrb}
)_{MF}=[B_{\alpha
\beta}^{*}(\vx-\vy) \mathbf{B}_{\vx+\vra,\vy+\vrb}-
A_{\alpha \beta}^{*}(\vx-\vy) \mathbf{A}_{\vx+\vra,\vy+\vrb}+\text{H.c}]
- \langle (\vec{\mathbf{S}}_{\vx+\vra}\cdot \vec{\mathbf{S}}_{\vy+\vrb} )_{MF}
\rangle \ea
\end{widetext}
where
\ba
\label{eq:mfeqs}
A_{\alpha \beta}^{*}(\vx-\vy)&=&\langle
\mathbf{A}^{\dag}_{\vx+\vra,\vy+\vrb}\rangle\\
B_{\alpha \beta}^{*}(\vx-\vy)&=&\langle \mathbf{B}^{\dag}_{\vx+\vra,\vy+\vrb}\rangle\nn
\ea
\ba
\langle (\vec{\mathbf{S}}_{\vx+\vra}\cdot \vec{\mathbf{S}}_{\vy+\vrb} )_{MF}
\rangle=|B_{\alpha\beta}(\vx-\vy)|^{2}-|A_{\alpha \beta}(\vx-\vy)|^{2}\nn\\
\ea
These are the mean field equations and must be solved in a self-consistent way together with the
constraints on the number of bosons
\ba
\label{eq:constraint}
B_{\alpha \alpha}(R=0)=2N_{c}S.
\ea
where $N_c$ is the total number of unit cells and $S$ is the spin strength. Numerical solutions of the (\ref{eq:mfeqs}) and (\ref{eq:constraint}) involves finding the roots of 24 coupled nonlinear equations for the parameters $A$ and $B$ plus the additional constraints to determine the values of the Lagrange multipliers $\lambda^{(\alpha)}$. We perform the calculations for finite but very large lattices and  finally we
extrapolate the results to the thermodynamic limit
\footnote{Generally the results for large systems (bigger than 2000 sites) did not show a
strong dependence on the size.}.

We solve numerically (\ref{eq:mfeqs}) and (\ref{eq:constraint}) for several values of
the frustration parameter $J_2/J_1$
and with the values obtained for the MF parameters and the Lagrange multipliers we
compute the energy and the new values for the MF parameters.
We repeat this procedure until the energy and the MF parameters converge.
After reaching convergence we can compute all physical
quantities like energy, spin-spin correlations and the excitation gap.
In order to support the analytical results of the MF approach,
we have also performed exact diagonalization on finite systems with 18, 24
and 32 spins with periodic boundary conditions for $S=1/2$ using Spinpack \cite{spinpack}.
\begin{figure}[t]
\includegraphics[width=0.35\textwidth]{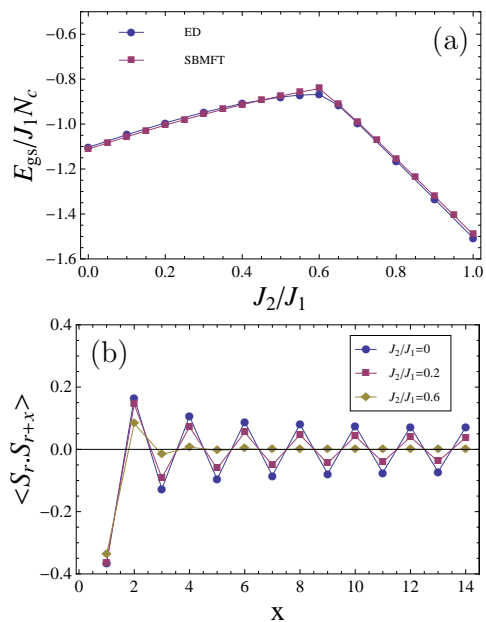}
\caption{(a): GS energy per unit cell as a function of $J_{2}/J_{1}$ for a lattice
of 32 sites. The circles are exact results (ED) and the squares are
the SBMFT results. (b): Spin-Spin correlation function (SSCF) vs distance $X$ in the {\it
zig-zag} direction obtained within SBMFT. For $0<J_2/J_1\lesssim0.41$, the SSCF correspond to the N\'eel phase white long-rage-order (LRO),  for $0.41\lesssim J_2/J_1\lesssim0.6$ the correlations are short ranged indicating a gap zone with sort-range-order (SRO).}
\label{fig:comparacion_E}
\end{figure}
In Fig. \ref{fig:comparacion_E}(a) we show the ground state energy per unit cell
as a  function of the frustration for a system of 32 sites calculated by mean
a SBMFT and ED showing an excellent agreement between both approaches. The advantage
of the SBMFT is that it allows to study much larger systems:
we have studied different system sizes up to 3200 sites, which we extrapolate
to the {\it thermodynamic limit}.

For the present model we only find commensurate collinear phases
(CCP) and for these phases the wave vector $Q_0$, where the dispersion
relation has a minimum, remains pinned at a commensurate point in the Brillouin
zone, independently of the value of the frustration $J_2/J_1$.
In the {\it thermodynamic limit} $N_{c}\to \infty$, a state with LRO is
characterized in the Schwinger boson approach by a condensation of bosons at
the wave vector $Q_0$. This implies that the
dispersion of the bosons in a state with LRO is gapless. As we discussed
earlier, we solve (\ref{eq:mfeqs}) and (\ref{eq:constraint}) for finite systems, then to detect LRO we
calculate the gap in the bosonic spectrum as a function of $J_2/J_1$  for
different system sizes and perform a finite size scaling finding a finite region
where the system remains gapped.
\begin{figure}[t]
\includegraphics[width=0.45\textwidth]{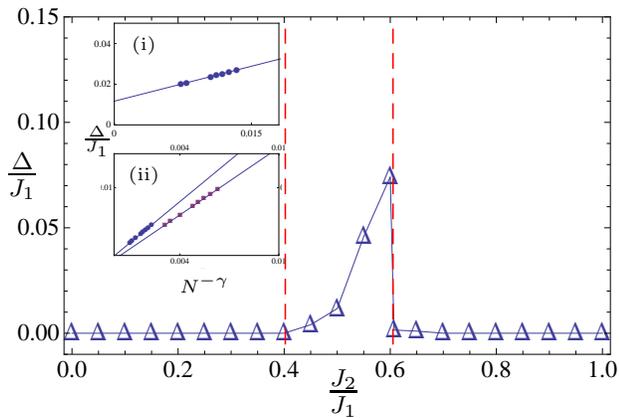}
   \caption{(a): Gap in the boson dispersion as a function of $J_{2}/J_{1}$
for $S=1/2$. In the region $J_{2}/J_{1}\sim 0.6$ the system remains gapped. Inset: finite size
scaling for the gap. (i) $J_2/J_1=0.5$ ($\gamma=0.6451$), (ii): Circles correspond to
$J_2/J_1=0.05$ ($\gamma=0.911$) and squares correspond to $J_2/J_1=0.35$ ($\gamma=0.758$).
}
    \label{fig:gap800}
\end{figure}
The extrapolation of the boson gap as a function of the frustration is shown
in Fig. \ref{fig:gap800}.
In the range of $0.41< J_2/J_1<0.6$ the system presents a finite spinon gap.
The inset shows an example of the finite size scaling for different values of the frustration.
The structure of the different phases can be understood calculating the
spin-spin correlation function (SSCF).  For $J_2/J_1<0.41$ the SSCF is
antiferromagnetic in all directions and is long-ranged while for $0.6<J_2/J_1$
we have found
ferromagnetic LRO correlations in the zig-zag direction that correspond to the
CAF
phase. The most interesting result is in the intermediate region $0.41<
J_2/J_1<0.6$ where the results for the SSCF are consistent with the
presence of a gap. In this region we have found short-ranged antiferromagnetic
correlations. A plot of the SSCF for $J_2/J_1=0,0.2$ and $0.6$ obtained within SBMFT is presented in Fig. \ref{fig:comparacion_E}b.

  The Fig. \ref{fig:Sc_NAF} shows the ground state phase diagram as a funtion of $1/S$. On one hand,
for $1/S$ smaller than a critical value $1/Sc(J_2/J_1)$, the correlation function have LRO, characterized
by a condensation of bosons at the wave vector $Q_0$. On the other hand, when
$1/S$ greater than $1/Sc(J_2/J_1)$, the correlation functions have SRO
indicating quantum disorder.

In summary, the results obtained with the different techniques used in the present paper suggest
the existence of a region in the intermediate frustration regime
where the system does not show quantum magnetic order for $S=1/2$.
On the one hand, LSW analysis predicts that the N\'eel LRO region extends up to
$J_2/J_1 \approx 0.29$ and the collinear antiferromagnet LRO  is present for
$J_2/J_1\gtrsim 0.55$. In the intermediate region we find no evidence of any
type of
magnetic order with this technique. On the other hand the results found with
SBMFT predict a quantum disordered region
$0.41\lesssim J_{2}/J_{1}\lesssim0.6$. In this region a gap opens in the bosonic dispersion and the
analysis of the spin-spin correlation function shows N\'eel short range order
followed by the LRO CAF phase for $J_2/J_1\gtrsim0.6$.

Acknowledgements: We would like to thank  G.L.\ Rossini, A.E.\ Trumper and L.O.\ Manuel for
helpful discussions and P.\ Pujol for pointing us an error in a previous version. This
work was partially supported by the ESF grant INSTANS, PICT ANPCYT (Grant No 20350) and PIP CONICET
(Grant No 1691).



\begin{thebibliography}{}
\bibitem{aclaracion} The $J_1-J_2-J_3$ Heisenberg model on the honeycomb lattice was analyzed previously using
LSWT in \cite{Fouet}, but the intermediate phase along the line $J_2 = J_3$, which is the main aim of the present
paper, was not analyzed in detail.
\end{thebibliography}
\end{document}